\documentclass{ws-procs9x6-arx}
\usepackage{amssymb}
\usepackage{graphicx}
\usepackage{epsfig}

\newcommand{\fs}{\; .}
\newcommand{\co}{\; ,}  
\newcommand{\non}{\nonumber}
\newcommand{\no}{\nonumber\\}
\newcommand{\cal}{\mathcal}
\newcommand{\br}{\langle}
\newcommand{\ke}{\rangle}
\newcommand{\unity}{{\bf 1}}
\newcommand{\mev}{{\rm MeV}}
\newcommand{\gev}{{\rm GeV}}
\newcommand{\nc}{{N_{\!c}}}
\newcommand{\eff}{{\rm eff}}
\newcommand{\QCD}{{\rm QCD}}
\newcommand{\atf}{^{\,\star}\hspace{-0.6em}}  
\newcommand{\btf}{^{\,\dagger}\hspace{-0.6em}}
\newcommand{\ctf}{^{\,\ddagger}\hspace{-0.6em}}

\newcommand{\dg}{\dagger}
\newcommand{\VR}{V_{\rm R}}
\newcommand{\VL}{V_{\rm L}}
\newcommand{\Tmat}{T}

\newcommand{\Fz}{F_0} %
\newcommand{\Bz}{B_0} %

\newcommand{\Meta}{M_{\eta}}
\newcommand{\Metap}{M_{\eta'}}
\newcommand{\Mz}{M_0}
\newcommand{\Sres}{{\cal S}}
\newcommand{\Pres}{{\cal P}}
\newcommand{\Ares}{{\cal A}}
\newcommand{\Vres}{{\cal V}}
\newcommand{\Rres}{{\cal R}}
\newcommand{\MS}{M_S}
\newcommand{\MP}{M_P}
\newcommand{\MV}{M_V}
\newcommand{\MA}{M_{\!A}}
\newcommand{\MR}{M_R}
\newcommand{\FV}{F_V}
\newcommand{\GV}{G_V}
\newcommand{\FA}{F_{\!A}}
\newcommand{\tJ}{J}
\newcommand{\JS}{\tJ_S}
\newcommand{\JP}{\tJ_P}
\newcommand{\JV}{\tJ_V}
\newcommand{\JA}{\tJ_{\!A}}
\newcommand{\JR}{\tJ_R}
\newcommand{\Hb}{H}
\newcommand{\Lb}{L}
\newcommand{\Ub}{U}
\newcommand{\dz}{d_0}

\newcommand{\tcL}{\tilde{\cal L}}
\newcommand{\tLeff}{\tilde{\cal L}_\eff} 
\newcommand{\tU}{\tilde{U}}
\newcommand{\tu}{\tilde{u}}
\newcommand{\tL}{\tilde{L}}
\newcommand{\tH}{\tilde{H}}
\newcommand{\tf}{\tilde{f}}
\newcommand{\tgamma}{\tilde{\gamma}}
\newcommand{\tchi}{\tilde{\chi}}

\newcommand{\tFz}{\tilde{F}_0} %
\newcommand{\tBz}{\tilde{B}_0}

\newcommand{\tpsi}{\tilde{\psi}}
\newcommand{\tLambda}{\tilde{\Lambda}}
\newcommand{\ttauz}{\tilde{\tau}_0}

\begin{document}

\title{Large $\nc$ in Chiral Resonance Lagrangians\footnote{{{\uppercase{UWT}h\uppercase{P}h-2004-42. \uppercase{W}ork supported by \uppercase{EC}-\uppercase{C}ontract \uppercase{HPRN}-\uppercase{CT}2002-00311 ({\uppercase{EURIDICE}}) and by \uppercase{A}cciones \uppercase{I}ntegradas, \uppercase{P}roject \uppercase{N}o. 19/2003 (\uppercase{A}ustria).}}}}

\author{ROLAND KAISER}

\address{Institute for Theoretical Physics, University of Vienna,\\ 
Boltzmanngasse 5,
A-1090 Vienna, AUSTRIA\\
E-mail: kaiser@ap.univie.ac.at}  

\maketitle

\abstracts{In the first part of the talk, I discussed results on the determination of the ratios of the light quark masses from large $\nc$ chiral perturbation theory, to be described elsewhere.\cite{KL-4}  The following notes contain material from the second part of the talk, which concerns the implications of large $\nc$ for resonance dominance estimates of the low energy coupling constants in chiral perturbation theory.}

\section{Introduction}

In the large~$\nc$ limit, $\nc \to \infty$, at fixed scale $\Lambda_\QCD$, the spectrum of QCD is known to consist of an infinite number of stable states.\cite{Large Nc,Manohar:1998xv,Lebed:2002tj}
The degrees of freedom in the corresponding low energy effective theory are the states with masses vanishing in the chiral limit ($m_u = m_d = m_s =0 $) of large~$\nc$ QCD, viz. $\pi$, $K$, $\eta$ and $\eta'$.\cite{Leff U(3)} The presence of the remaining massive states shows up only indirectly, trough their contributions  to the low energy expansion of QCD correlation functions. In chiral perturbation theory, these are accounted for in the form of low energy constants that arise at nonleading order in the low energy Lagrangian. 

In this note we discuss consequences for the low energy constants in the framework of an explicit realization of this scenario where, however, only the lowest lying resonance states are retained.   
In Sec.~\ref{LargeNChPT}, we briefly review the framework of the low energy expansion at large $\nc$. Secs.~\ref{Matter fields} and \ref{Resonance Lagrangians} deal with the accommodation of explicit resonance degrees of freedom in that setting.  Sec.~\ref{Constraints from QCD asymptotic behaviour} reviews the role of constraints from QCD asymptotic behaviour in the determination of the parameters occurring in the chiral resonance Lagrangian. In Sec.~\ref{Implications for the standard framework}, we discuss the implications of our analysis for the standard framework of chiral perturbation theory where $\nc$ is not treated as large. The corresponding numerical analysis may be found in Sec.~\ref{Numerical results}. Finally, Sec.~\ref{Discussion and conclusions} contains a discussion of the results and our conclusions.  

\section{Low Energy Expansion at Large $\nc$}
\label{LargeNChPT}

If the number of colours is treated as large, the low energy effective Lagrangian for QCD involves 9 degrees of freedom. The field variables are collected in a unitary matrix $\tU(x) \in $ U(3) and the extra field shows up as the phase of the determinant of $\tU(x)$,
\begin{eqnarray}\label{psi}
\det \tU(x)  =  e^{i \tpsi(x)} \fs
\end{eqnarray}  
The bookkeeping can be simplified by introducing a counting parameter $\delta$, where powers of the momenta, quark masses and $1/\nc$ are weighted according to 
\begin{eqnarray}
p^2 \sim m \sim 1/\nc = O(\delta) \fs
\end{eqnarray}
The expansion of the effective Lagrangian starts with a term of order $\delta^0$, 
\begin{eqnarray}
\label{Leff 1} 
\tLeff = {\tcL}_0 + {\tcL}_1 + {\tcL}_2 + O(\delta^3)    \fs
\end{eqnarray} 
In addition, the terms in the Lagrangian are subject to the constraints
\begin{eqnarray}
\tLeff \ge O(p^0) \; \co\;\; \tLeff \le O(\nc)  \fs
\end{eqnarray} 
According to these rules, the leading order term in the effective Lagrangian involves terms of order $\nc  p^2 $ and $\nc^0  p^0$, while the term of order $\delta$ collects the $\nc  p^4$, $\nc^0  p^2$ and $1/\nc p^0$ contributions, etc. For a detailed account of these matters, we refer the reader to Ref.~\refcite{Kaiser:2000gs}. 

The leading order Lagrangian in this expansion takes the form,\cite{Leff U(3),Gasser:1984gg,Kaiser:2000gs}   
\begin{align} \label{L0} 
\tcL_0 = &\; \tfrac{1}{4} \tFz^2 \br D_\mu \tU^\dagger D^\mu \tU + \tU^\dagger \tchi + \tchi^\dagger \tU \ke  - \tfrac{1}{2} \ttauz (\tpsi+\theta)^2 \co 
\end{align}
where $\theta$ is the field conjugate to the winding number density, which on account of the U(1)$_A$-anomaly transforms in such a manner that the combination $\tpsi + \theta$ remains invariant under chiral transformations.   
The external vector ($v_\mu$), axial vector ($a_\mu$), scalar ($s$) and pseudoscalar ($p$) fields enter in the expression for the covariant derivative $D_\mu \tU$ and $\tchi$,
\begin{align}
D_\mu \tU=  &\;  \partial_\mu \tU-i\,(v_\mu+a_\mu)\hspace{0.06em} \tU +i\,
\tU (v_\mu-a_\mu) \fs \non \\
\tchi= &\; 2 \tBz (s+i p) \fs
\end{align}

For a suitable choice of the effective variables, i.e. in particular the matrix $\tU$ and $\tpsi +\theta$, the individual terms in the effective Lagrangian in Eq. (\ref{Leff 1}) obey what we shall call `canonical large $\nc$ counting rules': These state that terms with a single trace are of order $\nc$, while the occurrence of each additional trace reduces the order {of} the term by unity. These rules also apply to terms involving powers of the chiral invariant combination $\tpsi + \theta$ if these are counted like extra traces.\cite{Kaiser:2000gs} The effective Lagrangian in Eq.~(\ref{L0}) exclusively involves fields which are of $O(1)$ in the large $\nc$ limit. In this case, the rules immediately apply to the coefficients of the terms and we deduce the following large $\nc$ behaviour for the three coupling constants in the Lagrangian in Eq. (\ref{L0}),
\begin{align}
 \tFz   =O(\sqrt{\nc}) \; \co \;\;  \tBz,\ttauz = O(1) \fs
\end{align}
$\tFz$ is pion decay constant in the limit of zero $u$, $d$ and $s$ quark masses, $\tBz$ is related to the quark condensate in the same limit, $\tBz = -  \br 0|\bar{u}u|0 \ke_0/\tFz^2 $. In the limit $\nc \to \infty$, $\ttauz$ coincides with the topological susceptibility of the corresponding quarkless theory (Gluodynamics) and equips the $\eta'$ with a mass of order $1/\sqrt{\nc}$,
\begin{align}\label{Mz}
\Metap^2 = \Mz^2 + O(m) \; \co \;\; \Mz^2 = \frac{6 \ttauz}{\tFz^2} = O(1/\nc) \fs 
\end{align}

At order $\delta$ the effective Lagrangian involves 11 additional low energy constants and takes the form,\cite{Kaiser:2000gs}
\begin{align} \label{L1} 
\tcL_1 = & \; \tfrac{1}{12} \tFz^2  \{ \tLambda_1 D_\mu \tpsi D^\mu\tpsi
- \tLambda_2 \,i(\tpsi+\theta)
\langle \tU^\dagger\tchi- \tchi^\dagger \tU\rangle \}  + \tfrac{1}{12} \tH_0 D_\mu\theta D^\mu\theta \no 
& +  \tL_2\langle D_\mu \tU^\dagger D_\nu \tU 
D^\mu \tU^\dagger D^\nu \tU\rangle +(2\tL_2+\tL_3) \langle D_\mu \tU^\dagger D^\mu \tU 
D_\nu \tU^\dagger D^\nu \tU\rangle \no 
&+
\tL_5 \langle D_\mu \tU^\dagger D^\mu \tU (
\tU^\dagger \tchi  +\tchi^\dagger \tU )\rangle +
\tL_8\langle \tU^\dagger \tchi \tU^\dagger\tchi +\tchi^\dagger \tU \tchi^\dagger \tU\rangle
\\
& -i \tL_9\langle R_{\mu\nu} D^\mu \tU D^\nu \tU^\dagger+
L_{\mu\nu} D^\mu \tU^\dagger D^\nu \tU\rangle+\tL_{10} \langle
\tU^\dagger R_{\mu\nu}\tU L^{\mu\nu} \rangle \no 
& +\tH_1\langle R_{\mu\nu} R^{\mu\nu}+
L_{\mu\nu} L^{\mu\nu}\rangle+\tH_2\langle\tchi^\dagger \tchi\rangle
\fs\nonumber
\end{align}
The invariant derivatives $D_\mu\tpsi$, $D_\mu\theta$ and the field strength tensors $R_{\mu \nu }$ and $L_{\mu \nu }$ are defined by
\begin{align}
D_\mu\tpsi & = \partial_\mu\tpsi -2\langle a_\mu\rangle\co 
& D_\mu\theta & =\partial_\mu\theta +2\langle a_\mu\rangle\co  \\ 
R_{\mu \nu } & = \partial_\mu r_\nu - \partial_\nu r_\mu - i
[r_\mu,r_\nu] \co  
& L_{\mu \nu } & = \partial_\mu l_\nu - \partial_\nu l_\mu - i
[l_\mu,l_\nu]\co
\non 
\label{notation}
\end{align}
where $r_\mu = v_\mu + a_\mu$ and $l_\mu = v_\mu - a_\mu$. The somewhat queer naming scheme for the {}coupling constants is chosen so as to facilitate the comparison with the standard framework, see Sec.~\ref{Implications for the standard framework}.  

In accordance with the canonical large $\nc$ counting rules, the terms in the first line of Eq.~(\ref{L1}) are of order $ \nc^0 p^2 $ and the remaining terms are $O( \nc p^4) $, viz.
\begin{align}
\tLambda_1, \tLambda_2 = O(1/\nc)  \; \co\;\;  \tH_0 = O(1) \; \co\;\; \tL_i , \tH_1, \tH_2 = O(\nc)  \fs
\end{align}  
In the following we are going to show how to obtain estimates for those coupling constants on the basis of a chiral Lagrangian with resonance fields. 

\section{Matter Fields}
\label{Matter fields}

The chiral transformation law for the effective field $\tU(x)$ is at the heart of the construction of the effective Lagrangians in Eqs. (\ref{L0}) and (\ref{L1}),  
\begin{align}\label{Utrafo}
\tU(x)' = \VR (x)\tU(x) \VL(x)^\dg \fs
\end{align}
with $\VR (x),\VL(x) \in$ U(3).
For the matter fields we need to find a corresponding transformation law, such that under transformations of the unbroken symmetry group $\VR(x)=\VL(x)=V(x)$ it reduces to the proper transformation law,\cite{Coleman:1969sm} e.g.
\begin{align}\label{matter:vector}
\Rres(x) ' = V(x) \Rres(x) V(x)^\dg \co
\end{align}
for a $3 \times 3$ matrix $\Rres$ collecting a nonet of resonance fields.  
To this end, introduce a unitary matrix $\tu(x)\in$ U(3), such that
\begin{align}\label{u:Def}
\tu(x)^2=\tU(x) \fs
\end{align} 
In order to promote Eq. ($\ref{u:Def}$) to a covariant relation, we deduce the following transformation law for the field $\tu(x)$,
\begin{align}
\tu(x)' = \VR (x)\tu(x) \Tmat(x)^\dg = \Tmat(x) \tu(x) \VL(x)^\dg \co
\end{align}
with a unitary matrix $\Tmat(x) \in$ U(3).
Note that for a general chi\-ral trans\-for\-mation, the matrix $\Tmat(x)$ depends not only on the transformation matrices $\VR(x)$ and $\VL(x)$ but also on the effective field $\tU(x)$, $\Tmat=\Tmat(\VR,\VL,\tU)$.\footnote{However, the matrix $T(x)$ hardly notices that we are considering unitary matrices: In fact, $T$ is independent of $\det \tU $ and $\det \VR \VL^\dagger$.} 
In the special case of a vector transformation $\VR(x)=\VL(x)=V(x)$, this dependence disappears, however: $\Tmat(x)=V(x)$. The transformation law
\begin{align}\label{matter:general}
\Rres(x) ' = \Tmat(x) \Rres(x) \Tmat(x)^\dg \co
\end{align}
does therefore represent one of possible extensions of the vector transformation law in Eq. (\ref{matter:vector}) to general chiral transformations. It furthermore has the advantage of  being right-left symmetric and preserving the hermiticity of the matter field $\Rres(x)$.

In this representation, the external fields $r_\mu $, $l_\mu$ and $\tchi$ appear in the following building blocks,\cite{Ecker:1988te}      
\begin{align}\label{lowercase fields}
\tu_\mu  & = \{ \tu^\dagger, i \partial_\mu \tu \} + \tu^\dagger r_\mu \tu - \tu\,    l_\mu \tu^\dagger   =  i  \tu^\dagger  D_\mu \tU \tu^\dagger 
\co \no 
\tf_{\pm\; \mu\nu}  & = \pm \tu^\dagger R_{\mu \nu} \tu +  \tu\,  L_{\mu \nu} \tu^\dagger
\co \\ 
\tchi_\pm &  =\tu^\dagger \tchi \,  \tu^\dagger \pm  \tu \, \tchi^\dagger \tu 
\co  \non
\end{align}
as well as in the covariant derivative associated with the transformation law in Eq. (\ref{matter:general}),
\begin{align}
\nabla_\mu \Rres & = \partial_{\mu} \Rres - i [\tgamma_\mu ,\Rres] 
\co \\
\tgamma_\mu & = \tfrac{1}{2} [ \tu^\dagger, i \partial_\mu \tu ] + \tfrac{1}{2} \tu^\dagger r_\mu \tu + \tfrac{1}{2} \tu\,   l_\mu \tu^\dagger 
\fs \non 
\end{align}
Expressed in terms of the lower case effective fields, the leading order Lagrangian in Eq. (\ref{L0}) reads
\begin{align}
\tcL_0 = \tfrac{1}{4} \tFz^2 \br \tu_\mu \tu^\mu +\tchi_+ \ke - \tfrac{1}{2} \ttauz (\tpsi +\theta)^2  \; \co \;\; \tpsi = - i \ln \det \tu^2  \fs
\end{align}
We are now in a position to proceed with the construction of the chiral Lagrangians for the resonance fields.

\section{Resonance Lagrangians}
\label{Resonance Lagrangians}

The chiral Lagrangians for vector ($\Vres$), axial vector ($\Ares$), scalar ($\Sres$) and pseudoscalar ($\Pres$) resonance fields take the form,\cite{Ecker:1988te} 
\begin{align}\label{LVA}
\tcL_R & =   -\tfrac{1}{2} \br  \nabla^{\lambda} \Rres_{\lambda\mu} \nabla_\nu  \Rres^{\nu\mu}  -  \tfrac{1}{2}\MR^2  \Rres_{\mu\nu} \Rres^{\mu\nu} \ke + \br \JR^{\mu\nu}  \Rres_{\mu\nu} \ke  \co 
&  \hspace{-0.5em}\Rres = \Vres, \Ares   \co \\ 
\label{LSP}
\tcL_R & =   \tfrac{1}{2} \br  \nabla_\mu \Rres \nabla^\mu  \Rres  -  \MR^2  \Rres^2  \ke + \br \JR  \Rres \ke  \co  
&\hspace{-0.5em} \Rres = \Sres, \Pres   \co
\end{align}
where, for convenience, the vector and axial vector resonances are described in terms of antisymmetric tensor fields, $ \Rres_{\nu\mu} =- \Rres_{\mu\nu}$.\cite{Gasser:1983yg,Ecker:1988te,Ecker:1989yg}
At order $\delta$, chiral symmetry permits the following set of independent contributions to the currents $\JR$,
\begin{align}
\JV^{\mu\nu} & = \tfrac{1}{2\sqrt{2}}\,  \{  \FV \tf^{\mu\nu}_+  + i \GV [\tu^\mu, \tu^\nu] \}  
\co \no 
\JA^{\mu\nu}  & = \tfrac{1 }{2\sqrt{2}} \, \FA  \tf^{\mu\nu}_- 
\co  \\
\JS & =   c_m \,  \tchi_ +  + c_d \, \tu_\mu \tu^\mu 
\co \no 
\JP & = i d_m \tchi_- + d_0 (\tpsi+\theta) \unity 
\fs \non 
\end{align}
Compared to the standard framework studied in Ref.~\refcite{Ecker:1988te}, our resonance Lagrangian involves one genuinely new contribution, in the pseudoscalar current $\JP$. We have denoted the corresponding coupling constant by $d_0$, while otherwise we have borrowed the notation of Ref.~\refcite{Ecker:1988te}.   

In the normalization convention of Eqs. (\ref{LVA}) and (\ref{LSP}), the resonance fields must be booked as order $\sqrt{\nc}$. The kinetic terms are then of order $\nc$, in accordance with the canonical large $\nc$ counting rules stated in Sec.~\ref{LargeNChPT}. The coupling constants in the resonance Lagrangian exhibit the following large $\nc$ behaviour
\begin{align}
\FV, \,\GV, \,\FA,\,c_m, \,c_d,\,d_m = O(\sqrt{\nc}) \; \co \;\;  d_0  = O(1/\sqrt{\nc})  
\co  
\end{align}
so that the terms involving the currents $ \JR$ are of order $\nc$ as well, with the exception of the piece proportional to $d_0$ which is of order 1 so as to account for the occurrence of the factor $\tpsi + \theta$. Finally, the resonance masses $\MS, \MP, \MV, \MA  $ are of order 1 in the large $\nc$  limit.\footnote{The discussion simplifies somewhat for rescaled resonance fields $\Rres' = \Rres/F_0= O(1)$.}    

When those masses are treated as large, the resonances may be integrated out, with the result\cite{Ecker:1988te}
\begin{align}
 \sum_{R=S,P,V,A} \hspace{-1em} \tcL_R =   - \frac{ \br \JV{\!}_{\mu\nu} {\JV}^{\mu\nu} \ke}{\MV^2} -  \frac{\br \JA{}_{\mu\nu}\JA^{\mu\nu}\ke}{\MA^2} + \frac{\br \JS^2 \ke}{2\MS^2}  +  \frac{\br \JP^2\ke}{2\MP^2} 
+ O(\delta^2) \fs
\end{align}
By use of the relations in Eq. (\ref{lowercase fields}) the result may be cast in the form 
\begin{equation}\label{L0+LRes}
\tcL_0 + \sum_{R=S,P,V,A}  \hspace{-1em} \tcL_R = \tcL_0 + \frac{3 \dz^2 }{2\MP^2} (\tpsi+ \theta)^2 + \tcL_\delta^\Rres + O(\delta^2) \co 
\end{equation}
where $\tcL_1^\Rres$ stands for an expression of the general form of the Lagrangian $\tcL_1$ in Eq.~\ref{L1} with specific values of the coupling constants, namely 
\begin{align}\label{Lambda2P}
\tLambda_1^\Rres  = 0 \; \co \;\; \tLambda_2^\Pres & = - \frac{ 12 \dz d_m }{\tFz^2 \MP^2}  = O(1/\nc)\; \co \;\;  \tH_0^\Rres  = 0  \co 
\end{align}
while the resonance contributions to the coupling constants $\tL_i$, $\tH_1$ and $\tH_2$ are all of order $\nc$ and listed in Table~\ref{Li-VASP}. 
Finally, the contribution proportional to $(\tpsi+ \theta)^2$ may be absorbed in an oder $1/\nc$ shift $ \ttauz \to \ttauz^\Pres $ according to 
\begin{align}
\ttauz^\Pres & = \ttauz - \frac{3 \dz^2 }{\MP^2} \fs 
\end{align}
Note that this correction has the right sign to explain why determinations of $\ttauz$ in the framework of lattice gauge theory\cite{tauGD Lattice,Hasenfratz:2002rp} would lead {to} values higher than those obtained from phenomenological determinations of the corresponding coupling constant $\ttauz^\Pres$ entering in the chiral Lagrangian. 
It is remarkable that the model fails to generate contributions to the coupling constants $\tLambda_1$ and $\tH_0$.   
With the phenomenological value $ \tLambda_2 - \frac{1}{2} \tLambda_1 \simeq 0.16 $,\cite{Leutwyler:1997yr,Diploma Thesis} we conclude that the product of the two coupling constants $d_0$ and $d_m$ is negative, $ d _0  d_m  < 0 $. In the following, we are going to adopt the convention $d_m > 0$.  

\begin{table}[t]
\tbl{Resonance contributions to the low energy coupling constants $\tL_i$ and $\tH_i$ arising at next to leading order in the framework of large $\nc$ chiral perturbation theory, cf. Eq. (\ref{L1}).}
{\begin{tabular}{ccccc}
\hline
{}&{}&{}&{}&{}\\[-.7em]
\hspace*{0.7em}&\hspace*{1.4em}$\Vres$\hspace*{1.4em}&\hspace*{1.4em}$\Ares$\hspace*{1.4em}&\hspace*{1.4em}$\Sres$\hspace*{1.4em}&\hspace*{1.4em}$\Pres$\hspace*{1.4em}\\[0.5em]
\hline 
{}&{}&{}&{}&{}\\[-.7em]
$\tL_2$			&$ \displaystyle \frac{\GV^2}{4\MV^2} $		&-								&-								&-									\\[1em]
$\tL_3$			&$ \displaystyle-3 \tL^\Vres_{2} $			&-								&$ \displaystyle\frac{c_d^2}{2\MS^2} $	&-									\\[1em]
$\tL_5$			&-									&-								&$ \displaystyle\frac{c_dc_m}{\MS^2} $	&-									\\[1em]
$\tL_8$			&-									&-								&$ \displaystyle\frac{c_m^2}{2\MS^2} $	&$\displaystyle- \frac{d_m^2}{2\MP^2} $		\\[1em]
$\tL_9$			&$ \displaystyle\frac{\FV \GV}{2\MV^2} $		&-								&-								&-									\\[1em]
$\tL_{10}$			&$ \displaystyle-\frac{\FV^2}{4\MV^2} $		&$\displaystyle\frac{\FA^2}{4\MA^2} $	&-								&-									\\[1em]
\hline
{}&{}&{}&{}&{}\\[-.7em]
$\tH_{1}$			&$\frac{1}{2} \tL_{10}^\Vres$				&$ - \frac{1}{2} \tL_{10}^\Ares $			&-								&-									\\[1em]
$\tH_{2}$			&-									&-								&$ \displaystyle 2\tL_8^\Sres $			&$ \displaystyle -2 \tL_8^\Pres  $			\\[0.5em]
\hline
\end{tabular}\label{Li-VASP}}
\vspace*{1em}
\end{table}

\section{Constraints from QCD Asymptotic Behaviour}
\label{Constraints from QCD asymptotic behaviour}

A way to obtain values for the parameters entering the chiral resonance Lagrangian is to relate them to the observed properties of the lowest lying resonances in the spectrum of QCD. Instead, we prefer to fix a maximal number of those coupling constants by considering the constraints that follow by imposing the proper asymptotic behaviour for massless QCD.\cite{Ecker:1988te,Ecker:1989yg,Leutwyler:1989pn}\cdash\cite{Cirigliano:2004ue} For the vector and axial vector resonances two such constraints may be inferred by considering the Weinberg sum rules,\cite{Weinberg:1967kj}  
\begin{align}
\label{W1} \FV^2 & = \tFz^2 + \FA^2  
\co \\
\label{W2} \FV^2 \MV^2 & = \FA^2 \MA^2  
\co 
\end{align}
demanding the asymptotic fall-off of the pion vector form factor,  
\begin{align}
\label{VFF} \FV \GV = \tFz^2 \co 
\end{align}
and the axial form factor $G_A(t)$,\cite{Ecker:1989yg} 
\begin{align}
\label{GA} \FV^2  = 2 \FV \GV  \fs
\end{align}
The four preceding equations allow us to express the three coupling constants $\FV, \GV, \FA$ in terms of $\tFz$ and the axial vector meson mass in terms of $\MV$,
\begin{align}\label{AVconstraints}
\frac{\FV}{\sqrt{2}} = \sqrt{2}{\GV} = \FA = \tFz  \; \co \;\; \MA = \sqrt{2} \MV \co 
\end{align} 
where we adopted the conventions $\FV, \FA  > 0$.
Inspection of the results in Table~\ref{Li-VASP} shows that this entails the prediction of the coupling constants $ \tL_2$, $ \tL_9$, $ \tL_{10}$ and $\tH_1$ in terms of the ratio $\tFz/\MV$,\cite{Ecker:1989yg}
\begin{align}
\tL_2^\Vres = \frac{1}{4} \tL_9^\Vres = - \frac{1}{3} \tL_{10}^{\Vres\Ares} = - \frac{2}{5} \tH_1^{\Vres\Ares} = \frac{\tFz^2}{8 \MV^2} \fs
\end{align}

In the scalar and pseudoscalar sector, there exists a constraint analogous to the one following from the first Weinberg sum rule in Eq. (\ref{W1}),\cite{Moussallam:1994at}  
\begin{align}
\label{SP-Weinberg} c_m^2  =  \frac{\tFz^2}{8} +d_m^2  \co 
\end{align}
as well as one condition from the asymptotic fall-off of the scalar form factor of the pion,\cite{Leutwyler:1989pn}
\begin{align}
\label{SFF} c_m  c_d =  \frac{\tFz^2}{4}  \fs
\end{align}
For a discussion of the spin 0 counterparts of the relations in Eqs.~(\ref{W2}) and (\ref{GA}) we refer the reader to Refs.~\refcite{Golterman:1999au,Pich:2002xy}. We point out that due to the absence of an equation analogous to Eq.~(\ref{W2}), one is free to consider the case where the pseudoscalar contribution is absent, viz. $d_m = d_0 =0$. In this case, the above equations imply    
\begin{align}
c_m = \frac{c_d}{2} = \frac{\tFz}{2\sqrt{2}} \fs
\end{align}   
($c_m > 0$) and lead to the prediction of $2 \tL_2+ \tL_3$, $\tL_5$, $\tL_8$ and $\tH_2$ in terms of $\tFz/\MS$, 
\begin{align}
(3\tL_2 +  \tL_3)^{\Vres\Sres} 
= \tL_{5}^{\Sres} 
= 4 \tL_{8}^{\Sres} 
= 2 \tH_2^{\Sres}
= \frac{\tFz^2}{4 \MS^2} \fs
\end{align}
The prediction for $\tL_5$ remains put, also if contributions from the pseudoscalar resonances are allowed -- this coupling constant depends only on the product $c_m c_d$ which is fixed by Eq. (\ref{SFF}). However, in this case the predictions for $3 \tL_2+ \tL_3$, $\tL_8$ and $\tH_2$ are modified according to 
\begin{align}
(3\tL_2 +  \tL_3)^{\Vres\Sres} & =  \frac{\tFz^2}{4 \MS^2}\, / \left[1+ \frac{8 d_m^2}{\tFz^2}\right]   
\co \no  
\tL_8^{\Sres \Pres}  & =  \frac{\tFz^2}{16\MS^2} + \frac{d_m^2}{2} \left[ \frac{1}{\MS^2}-\frac{1}{\MP^2} \right] 
\co \\ 
\tH_2^{\Sres \Pres}  & =  \frac{\tFz^2}{8\MS^2} + {d_m^2} \left[ \frac{1}{\MS^2}+\frac{1}{\MP^2} \right]
\co \non
\end{align}
where $c_m$ and $c_d$ have been traded for $d_m$. In particular we find $ \tL_8^{\Sres \Pres} \ge  \tL_8^{\Sres } $ as long as $ \MP \ge \MS$, i.e. the contributions from the pseudoscalars tend to increase the value of $\tL_8$.    

Before turning to the discussion of the numerical implications of the above, let us translate the results obtained so far to the standard framework of chiral perturbation theory, where more independent information on the values of the low energy coupling constants is available. 

\section{Implications for the Standard Framework}
\label{Implications for the standard framework}

If the number of colours is not treated as large, the $\eta'$ does no longer play a particular role but is just another of the states which remain massive in the chiral limit. In this case, the framework set up in Ref.~\refcite{Gasser:1984gg} provides the adequate description. There, the low energy expansion proceeds in powers of the momenta and light quark masses alone,
\begin{align}
{\cal L}_\eff = {\cal L}_2 + {\cal L}_4 +O(p^6)   
\end{align}  
In the following, we are going to exploit the fact that this framework effectively emerges from the theory discussed previously: We only need to consider it in the particular corner of its domain of validity where the mass of the $\eta'$ is large in comparison to the momenta and the light quark masses, while still being small in comparison to the intrinsic scale of QCD.  

To perform the matching procedure it is convenient to explicitly display the dependence on the singlet field $\tpsi$ by introducing an effective field $\Ub(x) \in$ U(3) according to\cite{Kaiser:2000gs} 
\begin{align}
\Ub  = e^{-\frac{i}{3}(\tpsi  +\theta )} \tU\fs
\end{align} 
Because the combination $\tpsi  +\theta $ represents a chiral invariant, the field $\Ub$ transforms in the same manner as $\tU$ in Eq. (\ref{Utrafo}). By its definition and Eq.~(\ref{psi}), it is further subject to the constraint 
\begin{align}
\det \Ub  = e^{- i\theta } \co 
\end{align} 
and does therefore describe the desired 8 degrees of freedom. To further simplify the discussion, we now switch off the singlet parts of the external fields and set 
\begin{align}
\theta = \br a_\mu \ke = \br v_\mu \ke = 0  \fs
\end{align}
When treating the $\eta'$ mass $ \Mz$ (\ref{Mz}) as large in comparison to the momenta and quark masses, the solution of the equation of motion for the singlet field $\tpsi$ implies the relations     
\begin{align}
\tU  = \Ub \{ 1+ \tfrac{i}{3} \tpsi \}  + O(p^4)  \;  \co \;\;  \tpsi = -\frac{i \tFz^2}{12 \ttauz^\Pres} \br \Ub^\dagger \tchi - \tchi^\dagger \Ub \ke + O(p^4)     \co
\end{align}
and it is a simple matter to convince oneself that the Lagrangian in Eq. (\ref{L0}) reduces to the standard form\cite{Gasser:1984gg} 
\begin{align}
{\cal L}_2 &  =  \tfrac{1}{4} \Fz^2 \br  D_\mu \Ub^\dg D^\mu \Ub +\Ub^\dg \chi+ \chi^\dg \Ub  \ke \co  \quad \chi = 2 \Bz (s+i p) \co 
\end{align}
with coinciding couplings $\Fz = \tFz$ and $ \Bz = \tBz$, up to contributions of order $p^4$.  
  
If this procedure is applied to the Lagrangian in Eq.~(\ref{L0+LRes}), the $O(p^4)$ piece of the result is of the form of the fourth order Lagrangian in Ref.~\refcite{Gasser:1984gg}, 
\begin{align}\label{L4su4}
{\cal L}_4 & =  \sum_{i=1}^{10} \Lb_i^\Rres O_i + \sum_{i=1}^{2} \Hb_i^\Rres {C}_i   \fs \non 
\end{align}
with
\begin{align}
 O_1 	& =  \br  D_\mu \Ub^\dg  D^\mu \Ub \ke ^2  \co 
 & O_2 & 	= \br  D_\mu \Ub^\dg  D_\nu \Ub \ke \br  D^\mu \Ub^\dg  D^\nu \Ub \ke  \co \no
  O_3 	& = \br  (D_\mu \Ub^\dg  D^\mu \Ub)^2 \ke  \co 
&  O_4  & = \br  D_\mu \Ub^\dg  D^\mu \Ub \ke  \br  \Ub^\dg \chi+ \chi^\dg \Ub \ke   \co \no
O_5 	& =\br  D_\mu \Ub^\dg  D^\mu \Ub (\Ub^\dg \chi+ \chi^\dg \Ub)  \ke  \co  
\hspace{-3em} & O_6 & 	= \br  \Ub^\dg \chi+ \chi^\dg \Ub\ke ^2 \co \\	
 O_7 	& =\br  \Ub^\dg \chi- \chi^\dg \Ub\ke ^2 \co 		
& O_8 	 & =\br  \Ub^\dg \chi\Ub^\dg \chi+\chi^\dg \Ub\chi^\dg \Ub\ke \co \no 
O_9 	& =-i \br  R_{\mu \nu} D^\mu \Ub  D^\nu \Ub^\dg + L_{\mu \nu} D^\mu \Ub^\dg D^\nu \Ub \ke \co 
\hspace{-5em} & \hspace{1em} & \hspace{3em} O_{10}  = \br  \Ub^\dg R_{\mu \nu} \Ub L^{\mu \nu} \ke  \co \no
 C_{1} 	& = \br  R_{\mu \nu} R^{\mu \nu} + L_{\mu \nu} L^{\mu \nu} \ke  \co 
& C_2 & 	= \br  \chi^\dg \chi \ke  \co  \non 
\end{align}
and\footnote{Here, the superscript $\Rres$ is meant to indicate the sum over all resonance contributions. However, the relations (\ref{LRSU3}) also hold for the individual contributions, $\Rres = \Vres, \Ares, \Sres, \Pres$.} 
\begin{align}\label{LRSU3}
2 \Lb_1^\Rres = \Lb_2^\Rres & = \tL_2^\Rres \; \co \; \;  &\Lb_3^\Rres &= \tL_3^\Rres \; \co \; \;  & \Lb_4^\Rres &= 0 \; \co \; \; &\Lb_5^\Rres &= \tL_5^\Rres \co \no
\Lb_6^\Rres & = 0  \; \co \; \; & \Lb_8^\Rres & = \tL_8^\Rres  \; \co \;\; &\Lb_9^\Rres &= \tL_9^\Rres \; \co \; \; & \Lb_{10}^\Rres &= \tL_{10}^\Rres  \fs \\ 
\Hb_1^\Rres & = \tH_1^\Rres  \; \co \; \;  & \Hb_2^\Rres & = \tH_2^\Rres  \fs  &&&&& \non 
\end{align}
Note that the term $ \br D_\mu \Ub^\dagger D_\nu \Ub D^\mu \Ub^\dagger D^\nu \Ub \ke $ has been eliminated by virtue of the algebraic identity\cite{Gasser:1984gg}
\begin{align}
\br  D_\mu \Ub^\dg  D_\nu \Ub  D^\mu \Ub^\dg  D^\nu \Ub \ke  = \tfrac{1}{2} O_1 +O_2  -2 O_3 \fs  
\end{align}
valid for $  \br \Ub^\dg D_\mu  \Ub \ke =0 $ (recall that we switched off the singlet external fields). Finally, the 
contribution to the coupling constant $\Lb_7$ is given by\cite{Leutwyler:1997yr,Kaiser:1998ds,Kaiser:2000gs}
\begin{align}\label{L7}
\Lb_7^\Rres=  - \frac{ \tFz^4 (1+ \tLambda_2^\Pres)^2}{288 \, \ttauz^\Pres}  \fs 
\end{align}
More information on the relation of the coupling constants in the two versions of the theory may be found in Ref.~\refcite{Kaiser:2000gs}. In particular the contributions generated by chiral loops as well as the contributions arising from nonvanishing singlet external fields may be found there.  

\section{Numerical Results}
\label{Numerical results}

For the numerical evaluation we employ the values (in $\mev$ units), 
\begin{align}
\Fz = 92.4  \; \co \;\; \MV = 770  \; \co \;\; \MS = 980  \; \co \;\; \MP= 1300  \co 
\end{align}
with intentional similarities to $F_\pi$, $M_\rho$, $M_{a_0}$ and $M_{\pi'}$, respectively.\cite{Eidelman:2004wy} With Eq. (\ref{AVconstraints}), this leads to the following numerical predictions for the coupling constants $ \FV $, $\GV $, $ \FA $ and the axial vector meson mass $ \MA $ ($\mev$ units), 
\begin{align}
\FV = 130.7 \; \co \;\; \GV = 65.3 \; \co \;\; \ \FA = 92.4 \; \co \;\; \MA= {1089 \fs} 
\end{align} 
To study the importance of the contributions of the pseudoscalar resonances we vary $d_m$ and consider the three values $ d_m=\{ 0,1, 2 \} \cdot \Fz /\sqrt{8} $. The equations (\ref{SP-Weinberg}) and (\ref{SFF}), then imply the following numerical values for $c_m$, $c_d$ and $d_m$  ($\mev$ units),
\begin{align}\label{cdm-num}
c_m & = 32.7 \co &c_d & = 65.3 \co &d_m & = 0  \co \no
c_m & = 46.2\co &c_d & = 46.2 \co &d_m & = 32.7  \co \\
c_m & = 73.0 \co &c_d & = 29.2 \co &d_m & = 65.3   \fs \non
\end{align} 

According to Eq. (\ref{Lambda2P}), the model fails to generate a contribution of the type $\tLambda_2 $ in the absence of contributions from pseudoscalar  resonances and we therefore set $\tLambda_2^\Pres =0$ for $d_m=0$. Otherwise, we adopt the phenomenological value $\tLambda_2^\Pres \to \tLambda_2 - \tfrac{1}{2} \tLambda_1  = 0.16$.\cite{Leutwyler:1997yr,Diploma Thesis} This coupling constant generates a shift of formal order $\nc$ in the coupling constant  $\Lb_7$, viz.
\begin{align}
\Lb_7   =- \frac{\tFz^2}{48 \Mz^2} (1+2 \tLambda_2^\Pres ) + O(1)  \co
\end{align}     
where in the above expression we have eliminated $ \ttauz^\Pres$ in favour of the mass of the $\eta'$ in the chiral limit, 
\begin{align}
\Mz^2 = {6 \ttauz^\Pres}{\tFz^{-2}} \co 
\end{align} 
and the numbers given in Table~\ref{Li-numerics} correspond to our favoured central value\cite{Kaiser:2002sz} of $ \Mz = 900 \,\mev $.  

\begin{table}[t]
\tbl{Numerical values for the coupling constants $\Lb_i $ of the order $p^4$ chiral Lagrangian, in units of $10^{-3}$. Rows 1, 2 and 3 list values for the $\Lb_i^r(\mu)$ at the scale $\mu = 770 \,\mev$ ($M_\rho$) obtained on the basis of phenomenology\protect\cite{Gasser:1984gg,Bijnens:1994ie} and lattice calculations.\protect\cite{Aubin:2004fs} Row 4 displays the resonance estimates for the $\Lb_i$ obtained in Ref.  \protect\refcite{Ecker:1988te}. The last column shows the numbers obtained in the present work, for three different values of the parameter $d_m$. 
}
{\begin{tabular}{rrrrrr}
\hline
{}&{}&{}&{}&&\\[-.9em]
					& $\Lb_i^r(M_\rho)$\cite{Gasser:1984gg}&  $\Lb_i^r(M_\rho)$\cite{Bijnens:1994ie}  	&$\Lb_i^r(M_\rho)$\cite{Aubin:2004fs}  &$\Lb_i^\Rres$\cite{Ecker:1988te}	&$d_m=\{ 0,1, 2 \}\cdot  \Fz /\sqrt{8} $	\\[.3em]
\hline
{}&{}&{}&{}&&\\[-.7em]
$\Lb_1^r$ 				& $0.7 \pm  0.3\atf$			&$0.4 \pm 0.3$    				& 						&$0.6$				&$0.9$\\[.6em]
$\Lb_2^r$ 				& $1.3 \pm 0.7$			&$1.35 \pm 0.3$    				& 						&$1.2$				&$1.8$\\[.6em]
$\Lb_3$ 				& $-4.4\pm 2.5\atf$			&$-3.5 \pm 1.1$    				& 						&$-3.0$				&$\{ -3.2, -4.3,-5.0\} $	\\[.6em]
$\Lb_4^r$ 				& $-0.3\pm 0.5\atf$			&							& $-0.1\pm 0.4\btf$			&$0$				&$0$\\[.6em]
$\Lb_5^r$ 				& $1.4 \pm 0.5$			&  							& $1.1\pm 0.4\btf $			&$1.4\ctf$				&$2.2$\\[.6em]
$\Lb_6^r$ 				& $-0.2\pm 0.3\atf$			& 							& $0.2\pm 0.3\btf$			&$0$				&$0$\\[.6em]
$\Lb_7$ 				& $-0.4 \pm 0.15$			&   							&						&$-0.3$				&$\{-0.2,-0.3, -0.3\} $	\\[.6em]
$\Lb_8^r$ 				& $0.9 \pm 0.3$			&    							& $0.6\pm 0.2\btf$			&$0.9\ctf$				&$\{ 0.6, 0.8,1.5\} $	\\[.6em]
$\Lb_9^r$ 				& $6.9\pm 0.7$				&   							&					&$6.9\ctf$				&$7.2$\\[.6em]
$\Lb_{10}^r$ 		         & $-5.5\pm 0.7$			&    							& 					&$-6.0$				&$-5.4$\\[.6em]
\hline
\end{tabular}\label{Li-numerics}}
\begin{tabnote}
$^\star$ Large $\nc$ estimate, $^\dagger$ Our error estimate, $^\ddagger$ Input from Ref.~\refcite{Gasser:1984gg}, see text for further explanations.
\end{tabnote}
\vspace*{1em} 
\end{table}

\section{Discussion and Conclusions}
\label{Discussion and conclusions}

In Table~\ref{Li-numerics}, the results of the present investigation are compared to values of the low energy constants $\Lb_i^r(M_\rho)$ obtained on the basis of phenomenology,\cite{Gasser:1984gg,Bijnens:1994ie} lattice QCD\cite{Aubin:2004fs} and the resonance estimates obtained in Ref.~\refcite{Ecker:1988te}. The resonance estimates from Refs.~\refcite{Ecker:1989yg,Pich:2002xy} are omitted in the table since they practically coincide with our values for the choice $d_m = \Fz/\sqrt{8}$. Our table also neglects values for the $\Lb_i$ obtained on the basis of a 2 loop phenomenonological analysis.\cite{Amoros:2001cp} A well known and unavoidable weakness of the method discussed in the present work lies in the fact that it fails to account for the known scale dependence of the coupling constants $\Lb_i^r(\mu)$. It is believed that the estimates should nevertheless give reasonable results, if the comparison is carried out at a typical hadronic scale $ \mu = 500, \ldots, 1000 \, \mev$. For definiteness, we list the values for the scale $\mu = 770 \, \mev$ ($M_\rho$).  

We point out that the values listed in Ref.~\refcite{Gasser:1984gg} do not all follow strictly from phenomenology, but in the case of $ \Lb_1$, $ \Lb_3$, $ \Lb_4$ and $ \Lb_6$ also rely on theoretical large $\nc$ arguments\cite{Gasser:1984gg}. Meanwhile, those predictions have been neatly confirmed in Ref.~\refcite{Bijnens:1994ie} on the basis of $K_{e4}$ and $\pi \pi$ data, and on the lattice\cite{Aubin:2004fs} (The errors quoted in Table~\ref{Li-numerics} follow if the individual errors listed in Ref.~\refcite{Aubin:2004fs} are added in quadrature). In the present framework, the predictions
\begin{align}
2 \Lb_1 - \Lb_2 = \Lb_4 = \Lb_6 = 0 \co 
\end{align}   
are simply an algebraic consequence of the absence of multiple trace terms in the resonance Lagrangian in Eqs. (\ref{LVA}) and (\ref{LSP}). 

The coupling constants $\Lb_1$,  $\Lb_2$, $\Lb_9$, $\Lb_{10}$ gain contributions exclusively from the vector and axial vector resonances (in view of the considerations in Sec.~\ref{Constraints from QCD asymptotic behaviour}, those should indeed be viewed as one entity). The predictions exhibit an impressive agreement with the values from Ref.~\refcite{Gasser:1984gg} and to a lesser extent also with those from Ref. \refcite{Bijnens:1994ie}.\footnote{The authors of Ref.~\refcite{Bijnens:1994ie} actually list two different sets of values for the $\Lb_i$ for the combined $K_{e4}$ and $\pi\pi$ data, obtained with different representations of the relevant form factors. The quoted fits give the central values $ \{\Lb_1^r(M_\rho),\Lb_2^r(M_\rho),\Lb_3\} =  \{ 0.6,1.5,-3.3\} \cdot 10^{-3}$ (one-loop) as well as those displayed in Table~\ref{Li-numerics} (unitarized), with similar statistical errors.} We should clarify at this point that the difference between the results of Ref.~\refcite{Ecker:1988te} and the present investigation is easily traced back to a difference in the numerical value of $\FV$ adopted in that reference, 
\begin{align}
\FV \simeq 154\;  \mev \co
\end{align}
the value that follows from the observed $ \rho^0 \to e^+ e^- $ rate.\cite{Eidelman:2004wy}
Accordingly, the authors of Ref.~\refcite{Ecker:1988te} do not make use of the relation (\ref{GA}), which in fact is known to be subject to corrections.\cite{Cirigliano:2004ue} Otherwise, our results coincide with those of Ref.~\refcite{Ecker:1988te}, and, for that matter, also with Refs.~\refcite{Ecker:1989yg,Pich:2002xy} -- the antisymmetric tensor fields $\Vres_{\mu\nu}$ and  $\Ares_{\mu\nu}$ simply do not notice the presence of the additional singlet pseudoscalar field. 

The prediction for $\Lb_5$ represents a scalar counterpart to the one for $\Lb_9$, but is clearly seen to work less well, in particular when compared to the lattice value.\cite{Aubin:2004fs} An obvious difference is seen in the magnitude of the two coupling constants as well: In the present picture, this fact finds an explanation in the difference of the vector and scalar meson masses.
The authors of Ref.~\refcite{Cirigliano:2003yq} present theoretical arguments in favour of a scalar mass of the order of $1.5 \, \gev$, which would help to resolve the discrepancies for $\Lb_5$. In any case, it should be noted that the coupling constant $\Lb_5$ is known to possess a strong scale dependence\cite{Gasser:1984gg} and thus varies significantly over the range $\mu = 500, \ldots,1000\, \mev $. With the central value of Ref.~\refcite{Gasser:1984gg}, the particular value $2.2 \cdot 10^{-3}$ is reached for $\mu \simeq \Meta$.       
    
In the absence of contributions from the nonet of pseudoscalar resonances ($d_m = 0$), our formulas imply $ \Lb_8 = \tfrac{1}{4} \Lb_5 $ leading to a rather low value for $\Lb_8 $ which, however, is in good agreement with the value from the lattice.\cite{Aubin:2004fs} In view of the discrepancy with $\Lb_5$, the combination $ 2 \Lb_8 - \Lb_5 $ turns out significantly negative, however, to be compared with $ 2 \Lb_8^r(M_\rho) - \Lb_5^r(M_\rho) = (0.2 \pm 0.2) \cdot 10^{-3}$.\cite{Aubin:2004fs} When the contributions from the pseudoscalars are switched on, the situation improves in that respect. Note that this is no longer true when $\MS$ exceeds $\MP$, $\MS> \MP$, in which case the value of $\Lb_8$ instead decreases when $d_m$ is growing. In Ref.~\refcite{Ecker:1988te}, the problems encountered here were circumvented  because the couplings $c_m$ and $c_d$ were instead determined so as to reproduce the central values of  $ \Lb_5 $ and $\Lb_8 $ from Ref.~\refcite{Gasser:1984gg} (with $d_m = 0$) leading to
\begin{align}
c_m \simeq 42\,  \mev \; \co \;\; c_d \simeq 32 \, \mev  \co
\end{align}
which is of course possible, at the cost of the validity of the relations in Eqs.~(\ref{SP-Weinberg}) and (\ref{SFF}), cf. also Eq.~(\ref{cdm-num}). 

Though of formal order $\nc^2$, the prediction for the coupling constant $ \Lb_7$ is known not to be extraordinarily large -- neither is the $\eta'$ extraordinarily light. The contributions from the additional pseudoscalar nonet lead to an additional small negative shift in $\Lb_7$. Note that the model should better predict a rather decent value for this constant, because in {}$\Lb_7$ and, for that matter, also $\Lb_3$, there is no scale dependence to be blamed for the discrepancy. In the case of $\Lb_3$, which is dominated by the vector and axial vector contributions, the phenomenological values\cite{Gasser:1984gg,Bijnens:1994ie} are not conclusive about the need for extra pseudoscalar contributions.               

In summary, the resonance dominance estimates for the coupling constants $\Lb_i$ have been demonstrated to lead to a rather coherent picture, also when the implications from large $\nc$ are taken seriously from the beginning to the end. The model involves a remarkably low number of adjustable parameters, and phenomenology appears to be in favour of the inclusion  of the contributions from the pseudoscalar $\pi'$ nonet.


{\bf Note added:} After the submission of the original manuscript it was pointed out to me that the list of references did only constitute an incomplete account of the recent work on resonance Lagrangians and large $\nc$ QCD. The present note aims at improving the work at hand in this respect: Work closely related to the present article work is described also in Refs. \refcite{Donoghue:1988ed,Bijnens:2001ps,Bijnens:2003rc}. The large $\nc$ approximation to QCD was tested qualitatively and quantitatively in Refs. \refcite{Knecht:1997ts}--\refcite{Peris:2000tw}. These references in particular introduced the concept of the `Minimal hadronic ansatz' denoting the smallest set of states required to match the given short distance behaviour. Large $\nc$ methods were further successfully applied to a wide range of phenomena such as the decay of pseudoscalars into lepton pairs,\cite{Knecht:1999gb} the evaluation of electroweak contributions to the pion mass difference,\cite{Knecht:1998sp} $K_0-\bar{K}_0$ mixing,\cite{Peris:2000sw} the weak matrix elements $Q_7$ and $Q_8$,\cite{Knecht:2001bc,Friot:2004ba} rare kaon decays,\cite{Greynat:2003ja,Friot:2004yr} electroweak hadronic contributions to the muon ${g-2}$\cite{Knecht:2002hr} and the determination of $\epsilon'/\epsilon$.\cite{Hambye:2003cy} For reviews of the subject we refer the reader to Refs.~\refcite{DeRafael:2001zs}--\refcite{Peris:2003gw}.

In the meantime, the work announced in Ref. \refcite{SPP} has appeared.\cite{Cirigliano:2005xn}           


\section*{Acknowledgments}
These notes emerged as a by-product of an ongoing collaboration\cite{SPP} with V.~Cirigliano, G.~Ecker, M.~Eidem\"uller, A.~Pich and J.~Portoles {who} are thanked for sharing their insights. We thank {G.~Ecker and} J.~Schweizer for useful comments on the manuscript and the organizers for the invitation to the workshop. Work supported by EC-Contract HPRN-CT2002-00311 (EURIDICE) and by Acciones Integradas, Project No. 19/2003 (Austria).


\end{document}